\documentclass{jkas}

\def\year{2017} % publication year
\def\month{May} % publication month
\def\volume{00} % journal volume
\def\issue{0} % journal issue
\def\beginpage{1} % first page of article
\def\endpage{99} % last page of article
\def\received{February 30, 2014} % date paper was received by JKAS
\def\accepted{February 31, 2014} % date of acceptance
\date{Received \received; accepted \accepted}

\newcommand\ion[2]{{#1}\,{\sc #2}} % ions: \ion{C}{iv} = C IV

\title{
Escape of Resonantly Scattered Ly$\beta$ and  H$\alpha$ from Hot and Optically Thick Media}
\author[1]{Seok-Jun Chang}
\author[1]{Hee-Won Lee}
\author[2]{Sang-Hyeon Ahn}
\author[2]{Hogyu Lee}
\author[3]{Rodolfo Angeloni}
\author[4]{Tali Palma}
\author[5]{Francesco Di Mille}
\affil[1]{Department of Physics and Astronomy, Sejong University, Neungdong-ro 209, Seoul 05006, Korea; \email{hwlee@sejong.ac.kr}}
\affil[2]{Korea Astronomy and Space Science Institute, Deajeon, Korea}
\affil[3]{Departamento de Fisica y Astronomia, Universidad de La Serena, Av. J. Cisternas 1200 Norte, La Serena, Chile}
\affil[4]{
Observatorio Astron\'{o}mico, Universidad Nacional de Córdoba, Laprida 854, C\'{o}rdoba, Argentina}
\affil[5]{Las Campanas Observatory, Carnegie Observatories, Casilla 601, La Serena, Chile}

\def\corrauthor{
H.-W. Lee
}

\def\runningauthor{
Chang et al.
}

\def\runningtitle{
Formation of H$\alpha$ and Ly$\beta$ 
}

\def\keywords{
radiative transfer -- scattering -- line formation -- line profiles 
}

\def\abstracttext{
We investigate the escape of Ly$\beta$ from emission nebulae with a significant
population of excited hydrogen atoms
in the level $n=2$, rendering them optically thick in H$\alpha$.
The transfer of Ly$\beta$ line photons in these optically thick regions is
complicated by the presence of another scattering channel leading
to re-emission of H$\alpha$, alternating their identities between
Ly$\beta$ and  H$\alpha$. In this work, we
develop a Monte Carlo code to simulate the transfer of Ly$\beta$ line
photons incorporating the scattering channel into H$\alpha$. Both H$\alpha$ and
Ly$\beta$ lines are formed through diffusion in frequency space, where a line
photon enters the wing regime after a fairly large number of resonance scatterings
with hydrogen atoms. Various line profiles of H$\alpha$ and Ly$\beta$
emergent from our model nebulae are presented. It is argued that the electron temperature is
a critical parameter which controls the flux ratio of emergent Ly$\beta$ and H$\alpha$.
Specifically for $T=3 \times 10^4{\rm\ K}$ and H$\alpha$ line center optical depth $\tau_\alpha=10$, the number flux
ratio of emergent Ly$\beta$ and H$\alpha$ is $\sim 49$ percent, which is quite significant.
We propose that the leaking Ly$\beta$ can be an interesting source for the formation of H$\alpha$ wings observed in
many symbiotic stars and active galactic nuclei. Similar broad H$\alpha$ wings are also expected in Ly$\alpha$ emitting halos found in
the early universe, which can be potentially probed by the {\it James Webb Telescope} in the future.
}

\begin{document}
\jkashead %% set title, authors, abstract, etc.

%%%%%%%%%%%%%%%%%%%%%%%%%%%%%%%%%%%%%%%%%%%%%%%%%%%%%%%%%%%%%%%%%%%%%
%%% BEGIN MAIN TEXT HERE %%%%%%%%%%%%%%%%%%%%%%%%%%%%%%%%%%%%%%%%%%%%
%%%%%%%%%%%%%%%%%%%%%%%%%%%%%%%%%%%%%%%%%%%%%%%%%%%%%%%%%%%%%%%%%%%%%

\section{Introduction\label{sec:intro}}

H$\alpha$ is one of the strongest optical emission lines in emission nebulae
found in a variety of celestial objects including
star forming regions, planetary nebulae and supernova remnants. In particular, emission nebulae
in symbiotic stars and active galactic nuclei (hereafter AGN) are proposed
to be formed through photoionization from a strong far UV source. Symbiotic stars are believed to be binary systems
of a white dwarf and a mass losing giant \citep[e.g.][]{kenyon86}. 
The white dwarf component in symbiotic stars exhibits various activities
including erratic variability and X-ray emission \citep[e.g.][]{nunez16}, which is attributed
to accretion by gravitational capture of a fraction of material lost
by the giant companion.  AGN are powered by a supermassive black hole
with an accretion disk \citep[e.g.][]{netzer15}. Strong emission of H$\alpha$ is also
present in Ly$\alpha$ emitting objects in the early universe \citep{dijkstra14}.

Spectroscopy using
{\it International Ultraviolet Explorer (IUE)} and {\it Far Ultraviolet Spectroscopic Explorer (FUSE)} shows that symbiotic stars and AGN
also share prominent emission lines with a large range of ionization potentials
including O~VI and Mg~II \citep[e.g.][]{godon12,sion17}. High ionization lines are mainly formed in a shallow region 
on the side illuminated by the strong
photoionizing source, whereas  low ionization lines are formed in a deep region shielded from highly ionizing radiation.
Therefore, the co-existence of high and low ionization lines implies that the emission regions are highly
optically thick. 
Ly$\alpha$ emitters in the early universe exhibit quite extended Ly$\alpha$ halo with significant polarization ,
which is consistent with considerable scattering optical depth of Ly$\alpha$ \citep{yang11}.
One consequence of the presence of highly optically thick media is the anomalous flux
ratios of Balmer lines, which deviates from those expected from the case B recombination
theory.

When the $n=2$ population of hydrogen is significant,
the transfer of H$\alpha$ line photons can be complicated due to another de-excitation channel.
That is, an excited hydrogen atom may de-excite into the ground state, reemitting a Ly$\beta$ photon.
The alternation of their identity between H$\alpha$
and Ly$\beta$ requires one to treat the transfer of H$\alpha$ and Ly$\beta$
together. In the case where  the level $n=2$ population is negligible compared
to the ground state population, Ly$\beta$ photons are selectively prevented
from escaping the region. The case B recombination theory is well established based on this
situation, in which Lyman photons are optically thick so that higher Lyman series
photons are absorbed by hydrogen atoms to reappear as Balmer or higher series photons.

The typical electron temperature of most emission nebulae is $10^4{\rm\ K}$ due to
balance between heating through ionization and cooling by line emission.
However, the thermal properties of emission nebulae in many symbiotic stars are consistent 
with electron temperature $T_e$ in the range of
 $10^4 - 4\times 10^4{\rm\ K}$ \citep[e.q.][]{skopal05}. In particular, \citet{sekeras12}
reported that broad wings around \ion{O}{6}~$\lambda\lambda$1032
and 1038 are consistent with the presence of Thomson scattering media with $T_e\sim 3\times 10^4{\rm\ K}$
in symbiotic stars.
Far UV observations of symbiotic stars show very strong resonance doublets
O~VI$\lambda\lambda$1032, 1038, N~V$\lambda\lambda$1238, 1243,
and C~IV$\lambda\lambda$1548, 1551, which are important coolants
of astrophysical plasma with temperature $T\sim 
10^5{\rm\ K}$ \citep[e.g.][]{gnat07}. O~VI lines are very strong
in symbiotic stars and quasars, implying that H$\alpha$ emission may also
be contributed in a significantly hot environment of $T\sim 10^5{\rm\ K}$.

With this temperature, the number of hydrogen atoms in an excited state
with $n=2$ is non-negligible compared to that in the ground state $n=1$.
In a medium that is optically thick at line center, a resonance line photon
escapes through diffusion in frequency space after a large
number of local scatterings. Due to the high optical depth for a photon with
wavelength near line center, it can escape only if it is
occasionally scattered off an atom with a large thermal motion along
the final line of sight. This last scattering allows the photon to enter
the wing regime, in which the medium is optically thin. This also implies
that the emergent resonance
line profile will exhibit a double-peak structure or equivalently will
be characterized by the presence of a central dip \citep[e.g.][]{neufeld90,gronke16}.

In the radiative transfer of optically thick H$\alpha$-Ly$\beta$,
diffusion in frequency space is also essential in a more complicated way.
The probability of making a de-excitation into $1s$ state is about 7 times
higher than into $2s$ state, so that we may expect that 
diffusion of Ly$\beta$ is more significant than H$\alpha$ in the Doppler factor space.
A Monte Carlo technique provides a straightforward method to treat the
transfer of resonance line photons, as is shown by many researchers.
In this paper, we develop a Monte Carlo code to deal with the radiative
transfer of H$\alpha$-Ly$\beta$ in optically thick media.

\section{Monte Carlo Approach to Radiative Transfer}

\subsection{Atomic Physics}

If an emission nebula is moderately optically thick at H$\alpha$ line center,
the transfer of H$\alpha$ photons mimics that of typical resonance line photons
in that escape is made through diffusion in frequency space \citep[e.g.][]{osterbrock62}.
In this respect, description of the transfer of H$\alpha$ in an optically
thick medium may start with the case of resonance line photons.
\citet{adams72} presented his early investigation of resonance line radiative
transfer using "Feautrier's method" to show that the mean number of
scatterings is almost proportional to the line center optical depth
of the medium. Diffusive nature in frequency space of the radiative transfer of resonantly
scattered photons in optically thick media
is beautifully illustrated in an analytical way by \citet{neufeld90}. 

The resonance line radiative transfer is also studied efficiently by adopting a Monte Carlo technique. In particular,
the transfer of Ly$\alpha$ is very important to understand the
physical properties of the intergalactic medium and star formation history
in the early universe \citep[e.g.,][]{dijkstra14, ouchi10, yang11}. A Monte Carlo approach turns out to be an efficient
method to compute the profile and polarization of resonantly scattered
Ly$\alpha$ \citep[e.g.,][]{ahn03,ahn15}.

A typical emission nebula around the hot white dwarf of symbiotic stars
has a physical dimension of $R\sim 10^{13}{\rm\ cm}$ and a proton
number density of $n_p \sim 10^{8}{\rm\ cm^{-3}}$. A neutral
fraction of $10^{-4}$ implies that H~I column density may reach
$N_{HI}\sim 10^{17}{\rm\ cm^{-2}}$, which is consistent with proposal
 that the emission nebula is ionization bounded.
Assuming as Boltzmann distribution with a temperature $T$, the
Ly$\beta$ line center optical depth is given by
\begin{equation}
\tau_\beta= N_{HI,1s}\sigma_{\nu_0}
\end{equation}
where $N_{HI,1s}$ is the column density of neutral hydrogen in the ground
state and $\sigma_{\nu_0}$ is the total cross section at Ly$\beta$ line center. For a typical
resonance line with a Gaussian line profile, the line center cross section is of order $10^{-13}T_4^{-1/2}{\rm\ cm^{2}}$,
where $T_4=T/10^4{\rm\ K}$ is the temperature in units of $10^4{\rm\ K}$ \citep[e.g.][]{rybicki85}.

The branching ratios of the two channels are obtained by considering the
spontaneous transition rates into $1s$ and $2s$ states from $3p$ state.
Following \citet{sakurai67} the spontaneous transition rate from state $A$ to state $B$ is
\begin{equation}
w_{BA}=\left({e^2\over 4\pi \hbar c}\right){4\over3} {\omega_{BA}^3\over c^2}|x_{BA}|^2,
\label{spon}
\end{equation}
where $\omega_{BA}$ and $|x_{BA}|$ are the angular frequency and matrix element of the
position operator between the states $A$ and $B$. The matrix elements are 
\begin{eqnarray}
|x_{1s,np}| &=&\left[ {2^8 n^7 (n-1)^{2n-5} \over 3(n+1)^{2n+5}} \right]^{1/2}a_B
\nonumber \\
|x_{2s,np}| &=&\left[ {2^{17} n^7 (n^2-1)(n-2)^{2n-6} \over 3(n+2)^{2n+6} }\right]^{1/2}a_B
\end{eqnarray}
where $a_B=\hbar^2/m_ee^2$ is the Bohr radius \citep[e.g.][]{saslow69}.

From Eq.~(\ref{spon} ) we obtain the ratio of transition probabilities from $3p$ to $2s$ and $1s$
\begin{equation}
{w_{3p\rightarrow2s}\over w_{3p\rightarrow1s}}={\omega_{23}^3\over\omega_{13}^3}
{|x_{2s,3p}|^2\over |x_{1s,3p}|^2}= \left({4\over5}\right)^9\simeq0.1342.
\end{equation}
Another way to look at this relation in terms of the oscillator strengths $f_{1s,3p}$ and $f_{2s,3p}$ is
\begin{equation}
{w_{3p\rightarrow2s}\over w_{3p\rightarrow1s}}={\omega_{23}f_{2s,3p}\over\omega_{13}f_{1s,3p}}=\left({4\over5}\right)^9
\end{equation}
where the numerical values of the oscillator strengths are
$f_{1s,3p}=0.07910, \quad f_{2s,3p}=0.4349$ \citep[e.g.][]{rybicki85}.
This implies that out of 8 scattering events about one transition is made into
the $2s$ state with an emission of H$\alpha$
\citep[e.g.][]{chang15, lee13}.

The line center optical depth $\tau_\alpha$
of H$\alpha$ due to hydrogen atoms in the $2s$ state is then related to $\tau_\beta$ by
\begin{equation}
\tau_\alpha = {4} e^{-\Delta E/kT}\tau_\beta,
\label{opt_depth}
\end{equation}
where $k$ is the Boltzmann constant and $\Delta E=10.2 {\rm\ eV}$ is the energy
difference between the hydrogen $1s$ and $2s$ states.
For $T=3\times10^4{\rm\ K}$, the ratio $\tau_\alpha/\tau_\beta\simeq 0.0663$.
In this work, the physical radius of a spherical emission nebula is measured by the line center optical depth.

H$\alpha$ line photons also arise from $3s-2p$ and $3d-3p$ transitions. 
These $3s-2p$  and $3d-2p$ transitions are meaningful only for initial generation of an H $\alpha$ photon that may directly escape. 
This is because if it is absorbed by another hydrogen atom in the $2p$ state, then effectively there is no generation of an $H\alpha$ photon 
from $3s-2p$ and/or $3d-2p$ transitions. If it is scattered by a hydrogen atom in the $2s$ state, then it is an identical process as we inject 
an H$\alpha$ photon generated from $3p-2s$ transition. In an optically thick medium, directly escaping fraction will be limited to those generated
near the surface, which may be negligible and disregarded in this work

\subsection{Monte Carlo Approach}

\begin{figure}
\includegraphics[width=80mm]{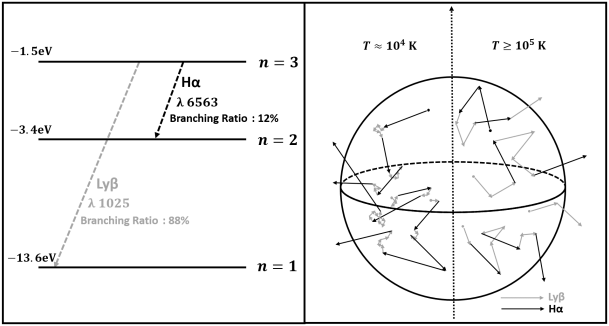}
\caption{Schematic illustration of radiative transfer of H$\alpha$ and Ly$\beta$ in an ionized
nebula.
The branching ratio into Ly$\beta$ transition from $3p$ is 0.88 and that into H$\alpha$ is 0.12.
As temperature goes up, the mean free path of Ly$\beta$ increases due to population shift to
$2s$ from $1s$.}
\label{scheme}
\end{figure}

\citet{ahn15} provided a detailed description of their Monte Carlo code,
which is also used in this work. 
Fig.~\ref{scheme} illustrates the mechanism of radiative transfer of H$\alpha$ and
Ly$\beta$ in an emission nebula. The $1s$ population being larger than $2s$ that of, implies that the line center optical
depth Ly$\beta$ of is larger than that of H$\alpha$, requiring higher scattering numbers for Ly$\beta$
to escape from the nebula. Because the branching into Ly$\beta$
is about 7 times more probable than that into H$\alpha$, on average a typical Ly$\beta$ photon
suffers 8 scatterings locally before it changes its identity into H$\alpha$.

In this work, H$\alpha$ line photons are
generated uniformly in a sphere with the line center
optical depth $\tau_\alpha$ ranging from 1 to 100.
A schematic description of our code is as follows.
The simulation starts with a generation of an initial H$\alpha$ photon
at a random position $\bf r$ in the sphere characterized by $\tau_\alpha$
and $\tau_\beta$.
The initial wavevector $\bf \hat k$ of propagation is chosen from
an isotropic distribution, and a Gaussian random deviate
is used to assign the initial Doppler factor $v$ along its direction of
propagation. 

A uniform random number $X$ between 0 and 1 is generated to determine the
optical depth $\tau$ the photon is supposed to traverse by the relation
\begin{equation}
\tau = -\ln (1-X).
\end{equation}

This optical depth is converted to physical free path length $s$ by the
relation 
\begin{equation}
{s \over R}=  {\tau \over \tau_\alpha} \exp{\left (v^2 \over v_{th}^2 \right )},
\end{equation}
where $R$ is the radius
of the spherical emission region. Here, $v_{th} =\sqrt{2kT/m_H}$ is the
thermal speed of hydrogen, whose mass is $m_H$. 
If the free path ends up inside the sphere, then a new scattering site is generated where we choose
the wavevector of a scattered photon.

In the frame of the scattering atom, the incident and outgoing photons are line center photons in the case of resonance scattering.. Therefore, 
the velocity component $v_1$ along the incident wavevector of the emitting atom coincides with that of the receiving atom. In a Monte Carlo approach,
one needs to specify the Doppler factor of a new scattered photon along its propagation direction given the Doppler factor along the incident direction.
Defining the scattering plane spanned by the incident and outgoing unit wavevectors ${\bf\hat k}_i$ and ${\bf\hat k}_o$, we introduce  two unit vectors 
${\bf\hat e}_1={\bf\hat k}_i$ and 
${\bf\hat e}_2={\bf\hat k}_i\times ({\bf\hat k}_i\times{\bf\hat k}_f)/|{\bf\hat k}_i\times
({\bf\hat k}_i\times{\bf\hat k}_f)|$, which constitute an orthonormal basis of the scattering plane. In the scattering plane, we may decompose the velocity ${\bf v}$ of the scatterer along ${\bf\hat e}_1$ and ${\bf\hat e}_2$ so that
\begin{equation}
{\bf v}=v_1{\bf\hat e}_1+v_2{\bf\hat e}_2
\end{equation} 
We know that $v_1$ is determined from the incident photon and we choose $v_2$ using a Gaussian random deviate. Then the velocity component $v_{\rm sc}$
along the propagation direction of the scattered radiation becomes
\begin{equation}
v_{\rm sc}={\bf v}\cdot {\bf\hat k}_o = v_1 \cos\theta+v_2\sin\theta, 
\end{equation}
where $\cos\theta={\bf\hat k}_i \cdot {\bf\hat k}_o$. From this consideration, the old Doppler factor $v_1$ is renewed to become $v_{\rm sc}$.

We use an isotropic scattering phase function in this work for simplicity.
In the case of resonance scattering, the scattering phase function differs according to the
angular momentum quantum numbers of the two levels \citep[e.g.][]{lee94}. However, we
are interested in emission nebulae with sufficiently large line center optical depths 
so that multiple scattering effectively results in isotropic angular distribution for scattered radiation.

Another uniform random number $X'$ is generated and in case it is smaller
than the branching ratio for de-excition to 2s then the new photon is still an H$\alpha$ photon
and otherwise we have a Ly$\beta$ photon \citep[e.g.][]{chang15}. 
A given photon is traced until
it escapes from the emission region keeping its identity and the Doppler factor
along its propagation.

\begin{figure}
\includegraphics[width=80mm]{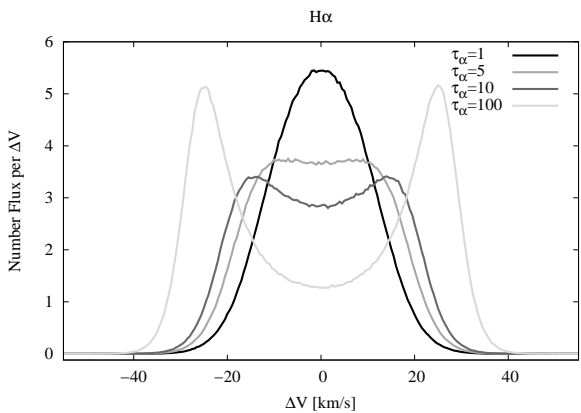}
\caption{Line profiles of H$\alpha$ emergent from a
scattering medium for various values of line center optical depth $\tau$
ranging from 1 to 100. No consideration of transition into the ground state
is made, treating H$\alpha$ as a genuine resonance line.}
\label{pure}
\end{figure}

First, no consideration of transition into the ground state is made  as a check of our Monte Carlo code.  
When only H$\alpha$ treated as a resonance line,  Fig.~\ref{pure} shows hypothetical profiles of H$\alpha$  emergent
from a spherical medium.  
 We collect all the photons escaping from the
emission region according to their Doppler factor or wavelength.
The temperature of the sphere is taken to be $T=10^4{\rm\ K}$ and line photons are generated 
with a Gaussian distribution. 
For $\tau_\alpha=1$, the effect of diffusion is negligible and the emergent profile
traces the Maxwell-Boltzmann distribution with the thermal speed $v_{th}=13{\rm\ km\ s^{-1}}$.
As $\tau$ increases, more and more line center photons
diffuse into the wing regime before they make their escape from the medium resulting in clear double
peak profiles. For $\tau_\alpha=10$, 100 and 1000,  the peaks are found to occur at $v=\pm 14{\rm\ km\ s^{-1}}$ and $22{\rm\ km\ s^{-1}}$
respectively.

\section{Line Transfer of H$\alpha$ and Ly$\beta$}

\begin{figure*}
\includegraphics[width=150mm]{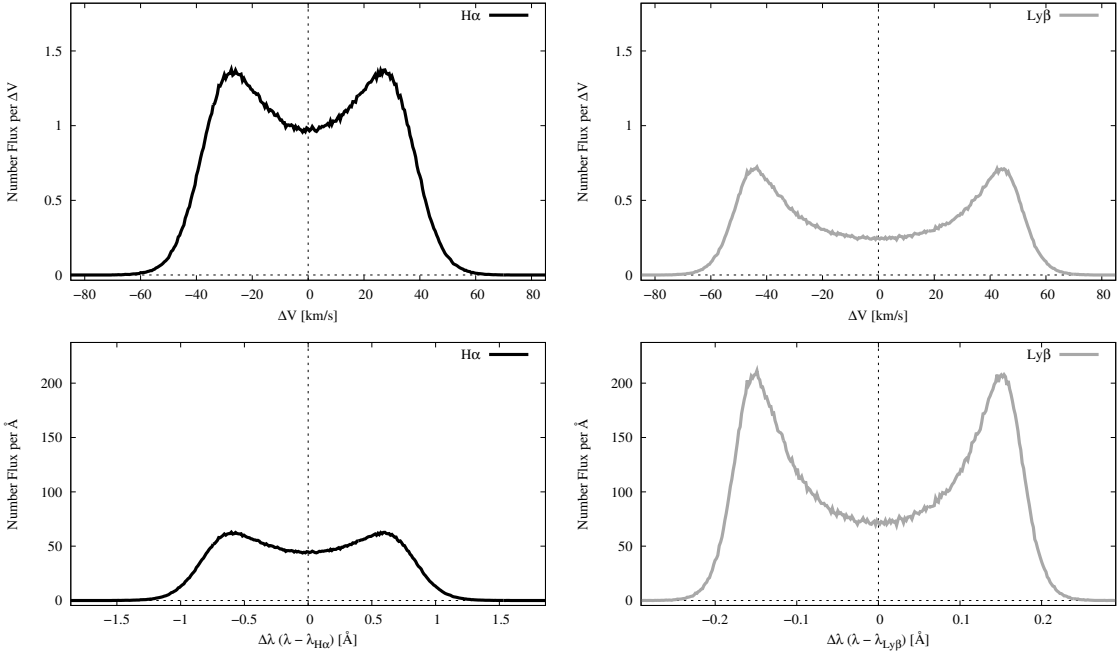}
\caption{Representative line profiles of emergent H$\alpha$ and Ly$\beta$ shown in
Doppler factor space (top panels) and wavelength space (bottom panels).
The emission nebula is characterized by the H$\alpha$ line center optical
depth $\tau_{\alpha}=10$ and temperature $T = 3\times10^4{\rm\ K}$.
It is noted that Ly$\beta$ exhibits wider profiles in Doppler space than H$\alpha$.
}
\label{doppler}
\end{figure*}

In Fig.~\ref{doppler}, we show a representative result from our Monte Carlo
calculation of H$\alpha$-Ly$\beta$ transfer. The spherical emission nebula
is characterized by $\tau_\alpha=10$ and $T=3\times 10^4{\rm\ K}$. The
two top panels show
line profiles of emergent H$\alpha$ and Ly$\beta$ shown in Doppler factor
space in units of ${\rm km\ s^{-1}}$. The vertical axis shows the number
flux density. The two bottom panels show the same result in wavelength space
in units of \AA.  In this particular example, the ratio of the emergent number fluxes
of Ly$\beta$ and H$\alpha$ is 0.486.

We obtain symmetric double peak profiles for both H$\alpha$ and
Ly$\beta$. In the Doppler factor space, the separation of the two peaks
is $80{\rm\ km\ s^{-1}}$ for H$\alpha$ and for Ly$\beta$
a slightly larger value of $105{\rm\ km\ s^{-1}}$ is obtained. 
We may also characterize the symmetric double peak profiles by the ratio of the number flux
density values at the peaks and the center dip, which are found to be 1.4 and 2.9 for H$\alpha$ and
Ly$\beta$, respectively.

\subsection{Dependence on Line Center Optical Depth and Temperature}\label{dep_tem}

\begin{figure*}
\includegraphics[width=160mm]{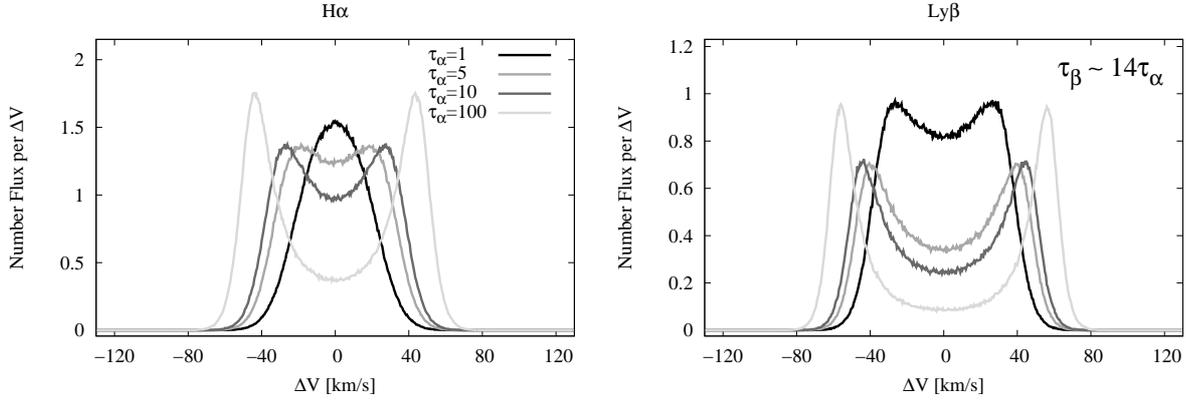}
\caption{Line profiles of H$\alpha$ (left) and Ly$\beta$ (right) for various values of H$\alpha$ line
center optical depth $\tau_\alpha$. Here, the nebular temperature is fixed $T = 3 \times 10^4 {\rm\ K}$.
}
\label{spec_depth}
\end{figure*}

In Fig.~\ref{spec_depth}, we show line profiles of H$\alpha$ and Ly$\beta$
by varying the line center optical depth $\tau_\alpha$ for a fixed value
of $T=3 \times 10^4{\rm\ K}$.  With this  $T = 3 \times 10^4{\rm \ K}$ it turns out that the ratio 
of optical depths $\tau_\beta / \tau_\alpha \sim 15$ so that these two transitions
have comparable scattering optical depths.
The vertical axis shows the number of
emergent photons per unit Doppler factor in units of ${\rm km\ s^{-1}}$.

As $\tau_\alpha$ increases, photons initially in the core part diffuse out
entering into the wing regime in frequency space, which results in line profile broadening.
In the Doppler factor space, the  Ly$\beta$ profile is always
broader than that of H$\alpha$, which is expected from the fact that
$\tau_\beta >\tau_\alpha$.
The profiles of emergent  Ly$\beta$ shown in this figure are doubly peaked through frequency diffusion. In contrast,
H$\alpha$ is singly peaked in the case of $\tau_\alpha=1$ for which frequency diffusion 
is minimal. As $\tau_\alpha \ge 5$ the profile flattens and begins to develop a clear double peak structure.
%The branching ratio to $1s$ state ($\sim 0.77$) is quite important parameter in relativly optically thin case to determine a flux ratio between Ly$\beta$ and H$\alpha$.
%The initial photons emitted near the surface are able to be emerged by singly scattering.
In this figure, the largest flux ratio $\sim 1.03$ of Ly$\beta$ to H$\alpha$ is obtained in the case of the smallest
scattering optical depth $\tau_\alpha=1$ considered in this work.
%We expect that a flux ratio between Ly$\beta$ and H$\alpha$ is less than unity and converges to the value multiplied by the branching ratio and the ratio of optical dpeths as long as $\tau_\alpha \gg 1$.

\begin{figure*}
\includegraphics[width=160mm]{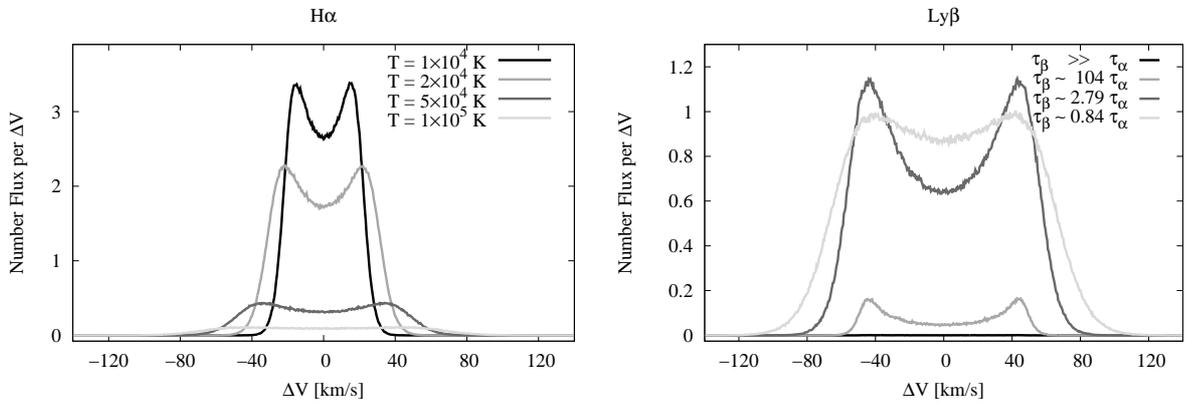}
\caption{Line profiles of H$\alpha$ (left) and Ly$\beta$ (right) for various values of $T$.
Here, the H$\alpha$ line center optical depth is fixed $\tau_\alpha=10$.
}
\label{spec_tem}
\end{figure*}

Fig.~\ref{spec_tem} shows various line profiles of emergent H$\alpha$ and Ly$\beta$
from a spherical emission region with a fixed value of $\tau_\alpha=10$. The temperature of the medium ranges
from $10^4{\rm\ K}$ to $10^5{\rm\ K}$.
As is verified in the figure, almost no Ly$\beta$ is observed in the case
$T=10^{4}{\rm\ K}$, justifying the use of the case B recombination theory in a typical emission nebula. 
However, as $T$ increases, the emergent Ly$\beta$
becomes strong and broad.

According to \citet{osterbrock62}, the mean scattering number of resonance
line photons before escape is linearly proportional to the line center
optical depth. This implies that the escape probability is inversely
related to the line center optical depth, which, in turn, leads to the relation
that the emergent line flux is also inversely proportional to
the line center optical depth. Because the ratio of the line center optical depths
of Ly$\beta$ and H$\alpha$ is directly related to the ratio of the $1s$ and
$2s$ populations, the ratio of emergent number flux is mainly determined by
the Boltzmann factor or equivalently the temperature of the medium.

\subsection{Flux Ratios in ($\tau-T$) Space}

\begin{figure*}
\includegraphics[width=170mm]{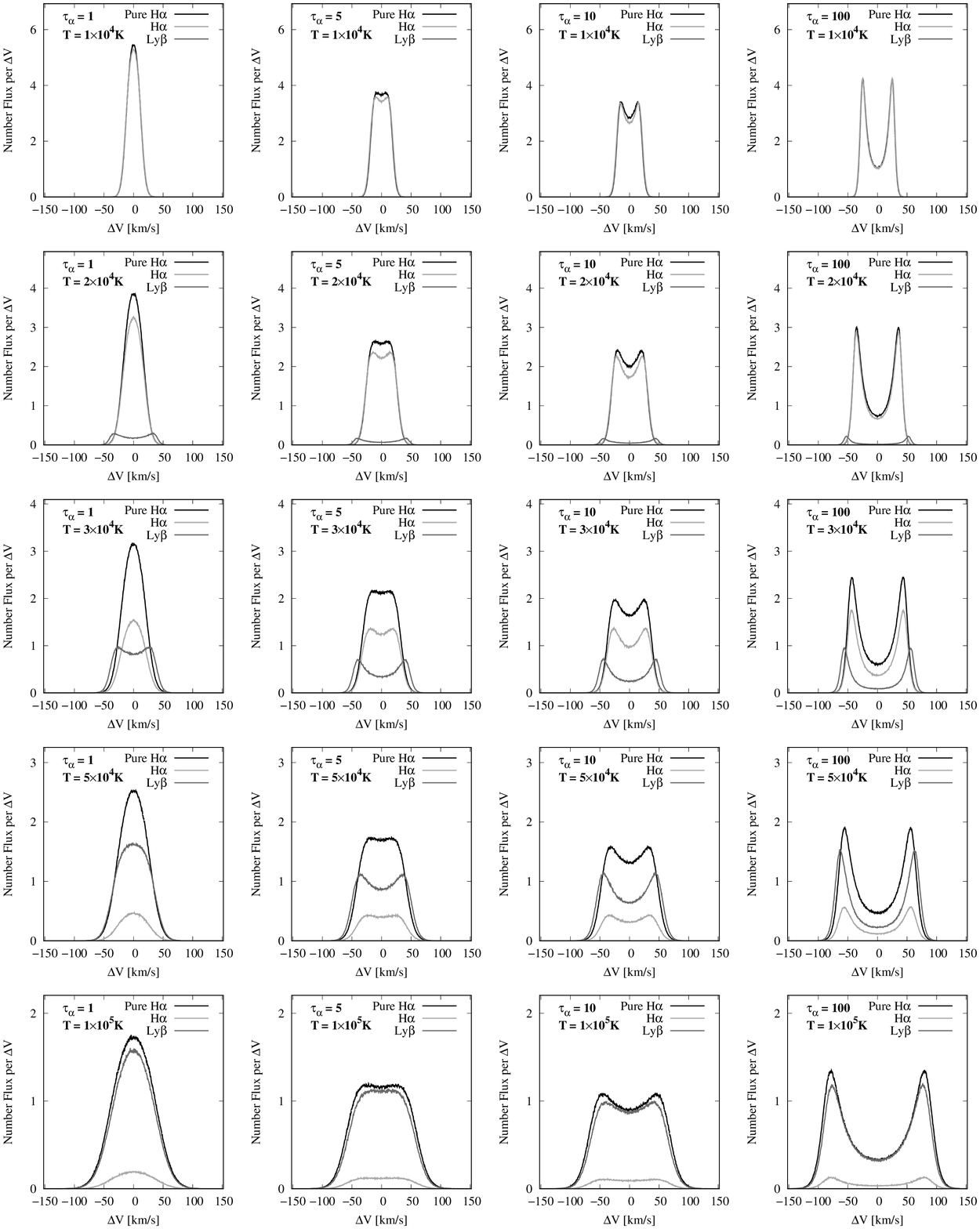}
\caption{Line profiles and number fluxes of emergent H$\alpha$ and Ly$\beta$
for various H$\alpha$ line center optical depth $\tau_\alpha$ and temperature $T$.
The ranges of  $T$  and $\tau_\alpha$ considered in this work are $10^4{\rm\ K} \le T \le 10^5{\rm K}$
and $1\le \tau_\alpha \le 100$ . The black curve is the emergent profile of H$\alpha$ with no consideration
of branching into Ly$\beta$ transition. Bright gray and dark gray lines show the profiles of H$\alpha$ and Ly$\beta$, respectively.}
\label{spec_multi}
\end{figure*}

%%%%%%%%%%%%%%%%%%%%%%%%%%%%%%%%%%%%%%%%%%%%%%%%%%%%%%%%%%%%%%%%%%%%%%%%%%%%%%%

In Fig.~\ref{spec_multi}, we present a suite of line profiles of
H$\alpha$ and Ly$\beta$ emergent from a spherical emission nebula
with a uniform density.  From left to right panels we increase
the line center optical depth $\tau_\alpha$ of H$\alpha$ and
from top to bottom panels the medium temperature increases from
$T=10^{4}{\rm\ K}$ to $T=10^5{\rm\ K}$.
The top-left panel for a small optical depth with $T=10^4{\rm\ K}$,
the emergent profile of H$\alpha$ is very close to the Gaussian function
corresponding to the Maxwell-Boltzmann distribution with the same temperature.
The profiles shown by a solid black line are the result that would be obtained if there is no
Ly$\beta$ scattering channel. Profiles of H$\alpha$ and Ly$\beta$ are shown by bright gray and 
dark gray solid lines, respectively. 

For $T= 10^4{\rm\ K}$, Ly$\beta$ is almost negligible, so that  H$\alpha$ behaves effectively
like a resonance line. This justifies use of the case B recombination theory, where no higher Lyman
series lines than Ly$\alpha$ are emergent with measurable fluxes. For $T=2\times 10^4{\rm\ K}$
a sizable fraction amounting $\sim 10$ percent of emergent line photons are Ly$\beta$. It is also
interesting to note that Ly$\beta$ exhibits double peak profiles for $T=2\times10^4, 3\times10^4{\rm\ K}$ and $\tau_\alpha=1$,
for which H$\alpha$ profiles are singly peaked.

\begin{figure}
\includegraphics[width=80mm]{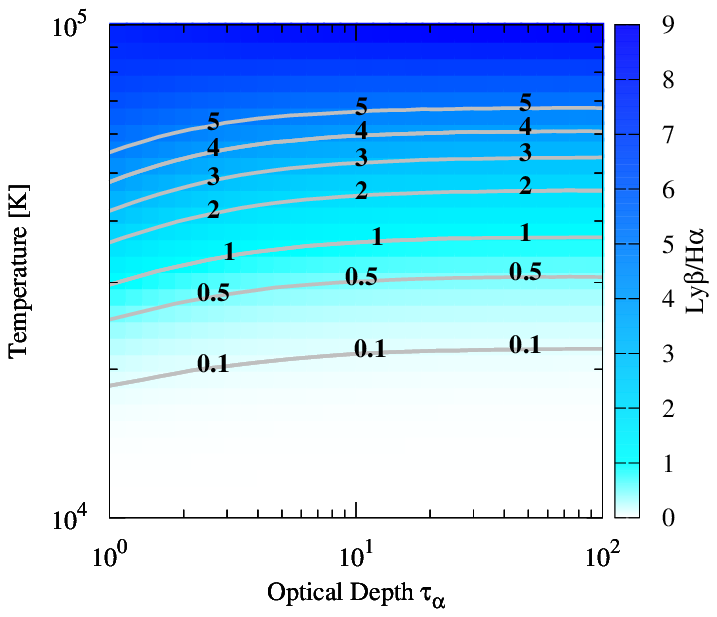}
\caption{Two dimensional plot of the number flux ratios of emergent H$\alpha$ and Ly$\beta$
for various values of H$\alpha$ line center optical depth $\tau_\alpha$ and temperature $T$.
As is discussed in the text, more Ly$\beta$ photons escape as $T$ increases and $\tau_\alpha$ decreases.
}
\label{flux_map}
\end{figure}

In Fig.~\ref{flux_map}, we present a color map of the ratio of the number fluxes of H$\alpha$ and Ly$\beta$ in the $(\tau_\alpha, T)$
space.  The contours are connecting those points of $(\tau_\alpha, T)$ yielding the same values of the number flux ratio.
The overall tendency is that Ly$\beta$ flux becomes relatively stronger as $T$ increases
and $\tau_\alpha$ decreases. It is readily seen that a negligible
flux of Ly$\beta$ is obtained in the case $T\sim 10^{4}{\rm\ K}$.
However, for $T=2 \times 10^4{\rm K}$ a considerably large ratio  of the number fluxes amounting to $\sim 0.1$  is obtained.
%The contour for the number flux ratio of 0.3 is quite flat in this figure indicating that the dominant
%controlling parameter is the medium temperature of $\sim 7\times 10^4{\rm\ K}$.  
When the medium temperature $T\ge 3 \times 10^4{\rm\ K}$
the number flux ratio exceeds 0.5, which is quite significant.

\subsection{Profile Widths of H$\alpha$ and Ly$\beta$}

\citet{osterbrock62} argued that the profile width of Ly$\alpha$ emerging from a medium with high scattering optical depth $\tau$ is
approximately proportional to $\sqrt{ \ln \tau}$. This expression results from the assumption that
resonance line photons escape after attaining frequency shift so that the medium eventually becomes optically thin.
Introducing a frequency parameter $x=(\nu-\nu_0)/\nu_{th}$ in units of the Doppler frequency width $\nu_{th}={v_{th}\over c}\nu_{0}$
with $\nu_0$ being line center frequency, the escape frequency $x_1$ will satisfy
\begin{equation}
\tau_\alpha e^{-x_1^2} \simeq 1,
\label{diffusion}
\end{equation}
leading to the dependence on $\sqrt{\ln\tau_\alpha}$. 

\begin{figure*}
\includegraphics[width=170mm]{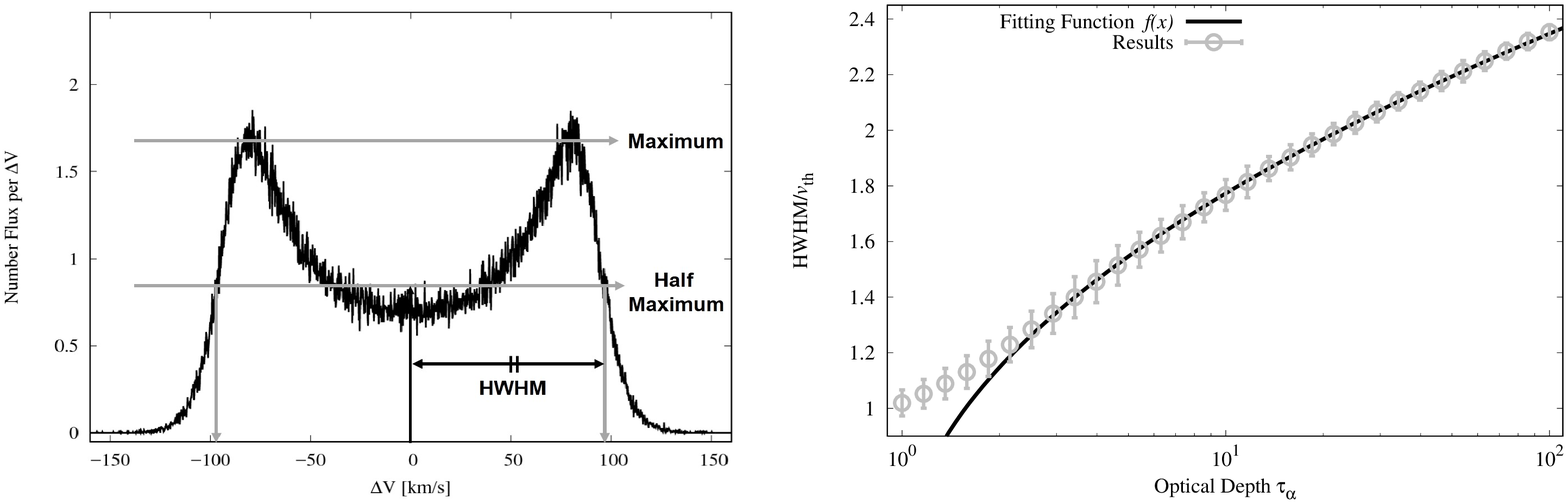}
\caption{Profile width as a function of H$\alpha$ line center optical depth. The left panel illustrates the definition of the half width at half maximum (HWHM).
The right panel shows ${\rm HWMH}$ of H$\alpha$ divided by the thermal speed $v_{th}$. We obtain the HWHM
by averaging the Monte Carlo data obtained for 31 values of
$T$ in the range of $10^{4}-10^{5}{\rm\ K}$ for a given value of $\tau_\alpha$. The error bars represent one standard deviation.  
The fitting function $f(\tau_\alpha)$ is also shown
by a solid curve, where $f(\tau_\alpha)=0.91(\ln \tau_\alpha)^{1/2} + 0.38$.
}
\label{fit}
\end{figure*}

 We quantify the diffusion process in frequency space by estimating
the half width at half maximum (HWHM) of line profiles.
In the right panel of Fig.~\ref{fit}, we show the method to measure the HWHM and the result of our Monte Carlo calculation.
Given $\tau_\alpha$ we consider 31 cases of $T$ in the range $10^{4-5} {\rm\ K}$ to obtain emergent H$\alpha$
profiles. The HWHM of H$\alpha$ is divided by $v_{th}$ in order to clarify the diffusive nature of line radiative transfer in frequency space.
The average and the standard deviation of the HWHM are subsequently obtained to be shown in the right
panel by circles with an error bar. The smallness of the error bars justifies this approach to present the profile width 
as a function $\tau_\alpha$.
 
The data can be fit with a function $f(\tau_\alpha)$
\begin{equation}
f(\tau_\alpha) = 0.91\sqrt{\ln \tau_\alpha} + 0.38.
\end{equation}
A good fit is obtained for $\tau_\alpha>3$, where Eq.~\ref{diffusion} is approximately valid. However,
for $\tau_\alpha < 3$, 
the optical depth is not high enough to make the radiative transfer diffusive in frequency space.
This is also confirmed from the single peak profiles exhibited by H$\alpha$ in this low optical depth regime.

\begin{figure*}
\includegraphics[width=170mm]{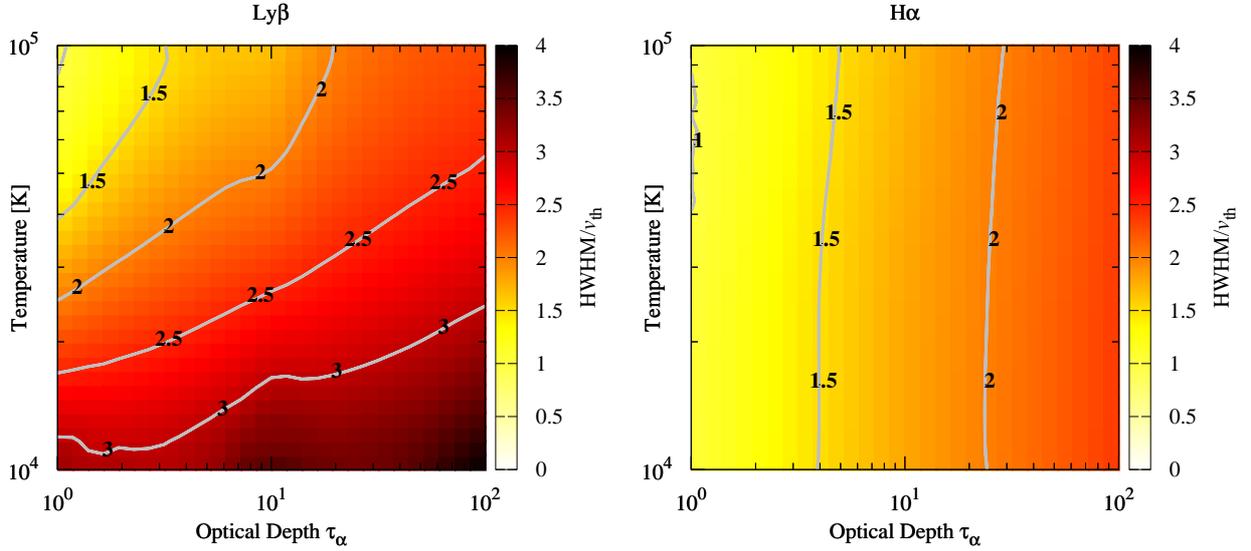}
\caption{Profile widths of Ly$\beta$ (left panel) and H$\alpha$ (right panel) 
normalized by the thermal velocity $v_{th}$ for various values of $\tau_\alpha$ and $T$
of the emission region. The horizontal axis is line center optical depth $\tau_\alpha$ and the vertical axis is
the temperature $T$ of the emission nebula.
}
\label{widths}
\end{figure*}

Fig.~\ref{widths} presents the HWHM of Ly$\beta$ and H$\alpha$ profiles  in units
of $v_{th}$ in the 2 dimensional parameter space of $\tau_\alpha$ and $T$.
In the case of H$\alpha$, the HWHM appears to depend only on line center optical depth $\tau_\alpha$ and is almost independent of $T$. 
This is attributed to normalization by $v_{th}$, which corresponds to the one-step interval as we approximate a diffusion phenomenon
by a random walk process.
Ly$\beta$ profile widths are very noisy for $T\sim10^4$ because of the small number statistics of escaping Ly$\beta$.
According to Eq.~\ref{opt_depth}, the Ly$\beta$ optical depth $\tau_\beta$ is larger than the H$\alpha$ counterpart $\tau_\alpha$,
which gives rise to broader Ly$\beta$ than H$\alpha$.

\subsection{Scattering Numbers of H$\alpha$ and Ly$\beta$}

\begin{figure*}
\centering
\includegraphics[width=175mm]{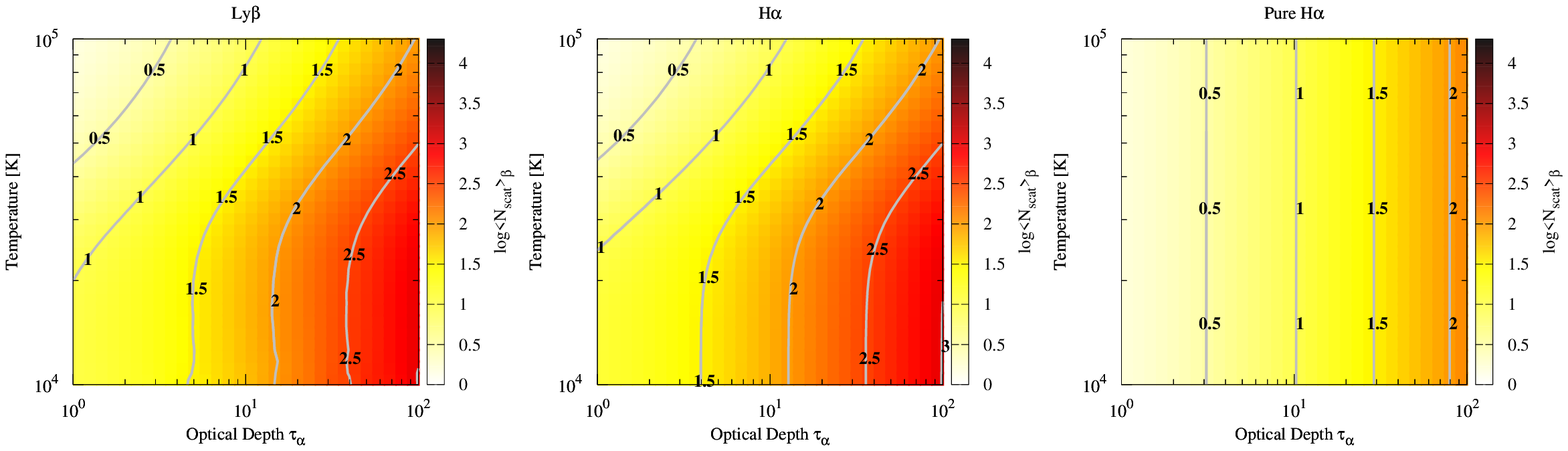}
\caption{Scattering numbers of H$\alpha$ and Ly$\beta$ before escape
for various $\tau_\alpha$ and the temperature
of the emission region. The horizontal axis is line center optical depth and the vertical axis is
the temperature of the emission nebula.
}
\label{ns}
\end{figure*}

In Fig.~\ref{ns}, we show the scattering numbers of Ly$\beta$ and H$\alpha$ before escape for
various $\tau_\alpha$ and the temperature of the emission region. For comparison in the
third panel we show the scattering number of H$\alpha$ treating it as a resonance line by setting the branching ratio
into Ly$\beta$ to be null.  In the case of resonance line
radiative transfer, the scattering number is approximately proportional to the line center optical 
depth \citep[e.g.,][]{adams72}. 

In the second panel, the scattering number of H$\alpha$ is slightly larger than that expected for resonance
line photons in a medium with the same line center optical depth, which is attributed to an additional
contribution from the Ly$\beta$ scattering channel. The contribution of the Ly$\beta$ channel increases
as the level $n=1$ population increases. This implies that given $\tau_\alpha$ the scattering number of emergent
H$\alpha$ decreases as $T$ increases.

In the first panel, the Monte Carlo data for $T\sim 10^4{\rm\ K}$ are noisy, which is attributed to
a very weak emergent Ly$\beta$ flux in these cases. Furthermore, the scattering number is rather small compared to the line center optical
depth. This is explained by the fact that the weak emergent Ly$\beta$ flux is significantly contributed by those line photons 
originating near the boundary. It is also noteworthy that as $T\ge 3 \times 10^4{\rm\ K}$ the scattering numbers of Ly$\beta$ and H$\alpha$ 
become comparable to each other.

It should be noted that the left and center panels show remarkably similar behavior of the scattering number.
If $\tau_\alpha$ and $\tau_\beta$ are both large, then the transfer processes of Ly beta and H alpha are virtually same and the final scattering 
determines the identity of the escaping photon. Therefore as far as $\tau_\alpha$ is larger than unity, we expect the virtually similar dependence 
of scattering number of H alpha and Ly beta on tau and temperature.

\section{Application to Broad H$\alpha$ Wings in Symbiotic Stars}

\begin{figure}
\centering
\includegraphics[width=85mm]{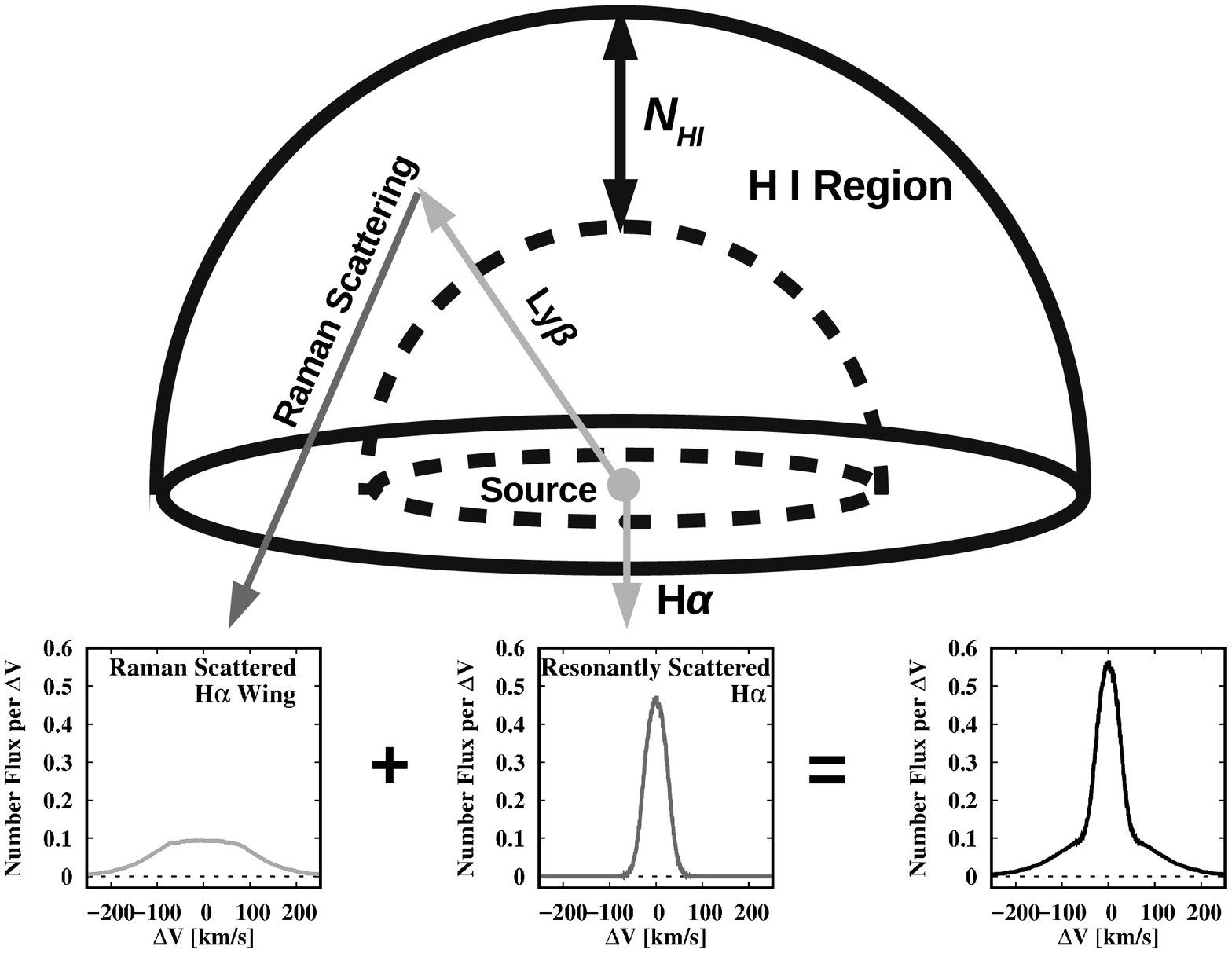}
\caption{Schematic illustration of forming H$\alpha$ wings in a hemispheric
neutral region surrounding a central H~II region through Raman scattering 
of Ly$\beta$.
The center emission source is a spherical ionized region shown in the right panel 
of Fig.~\ref{scheme}.
$N_{HI}$ is the H~I column density of hemispheric neutral region.
The center panel shows H$\alpha$ emission profile emergent from the source 
characterized with $T = 5 \times 10^4 \rm K$ and $\tau_\alpha = 1$.
The left panel shows Raman scattering wings obtained with $N_{HI}=10^{19} \rm cm^{-2}$ 
The right panel shows the composite profile of H$\alpha$ emission and wings.
}
\label{raman}
\end{figure}

\begin{figure*}
\centering
\includegraphics[width=175mm]{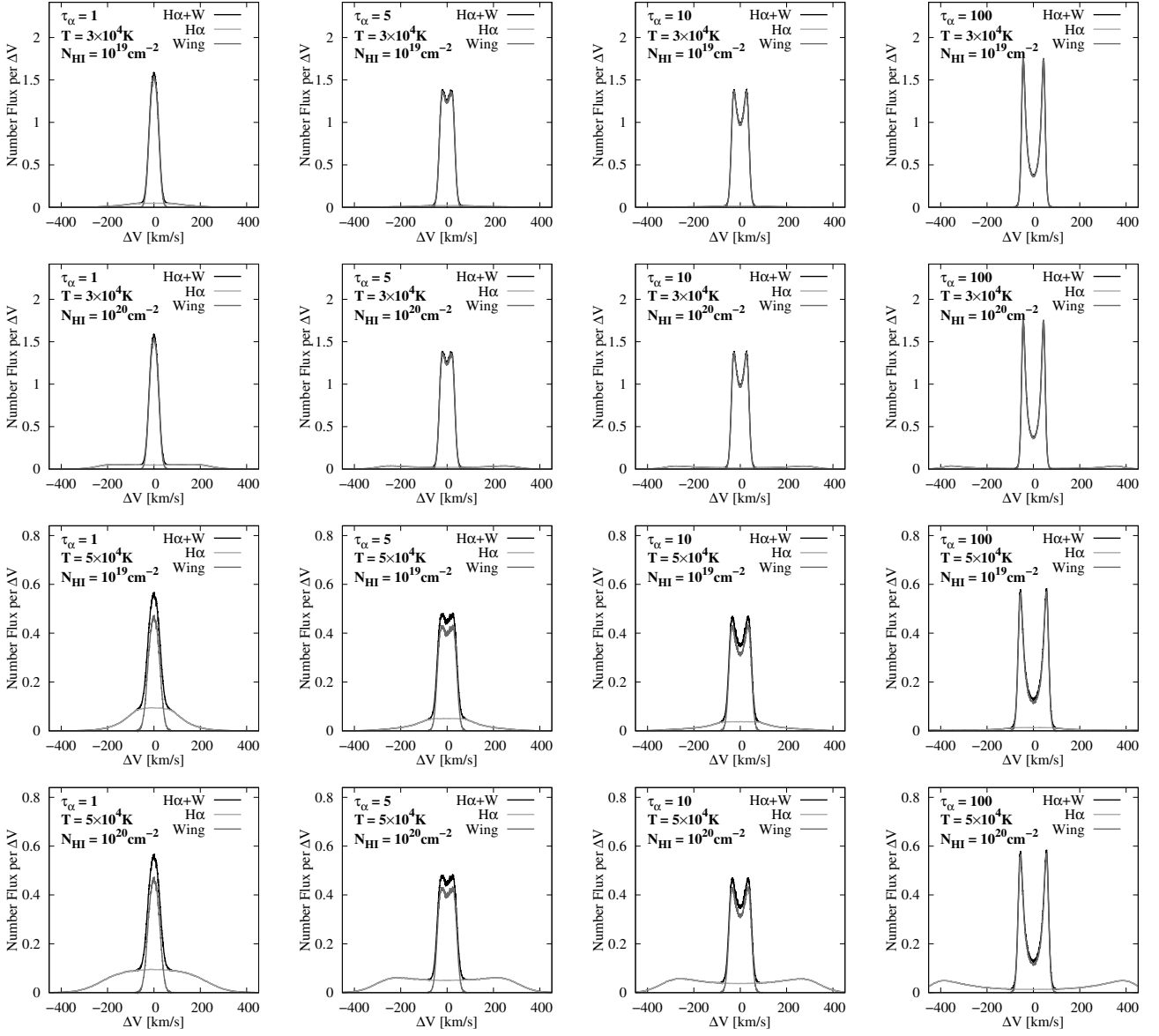}
\caption{Line profiles and number fluxes of Raman scattered H$\alpha$ wings (dark gray) and
H$\alpha$ emission (bright gray) in the model illustrated in Fig.~\ref{raman}.
The black solid lines show the total profiles combining the wings and emission.
The source temperature $T=3 \times 10^4 \, \rm K$ for the top two panels 
and $T=5 \times 10^4 \, \rm K$ for the bottom two panels. 
The optical depth $\tau_\alpha$ considered is in the range $1 \le \tau_\alpha \le 100$.
%The column densities of the neutral region $N_{HI}$ are $10^{19} \rm cm^{-2}$ for the first 
%and thrid panels and $10^{20} \rm cm^{-2}$ for the second and fourth panels.
Two values of $N_{HI}=10^{19}\, \rm cm^{-2}$ and $10^{20}\, \rm cm^{-2}$ are considered.
}
\label{raman_wing}
\end{figure*}

Spectroscopic observations of symbiotic stars reveal that H$\alpha$ emission lines
exhibit broad wings that may extend to $10^3{\rm\ km\ s^{-1}}$ \citep[e.g.][]{vanwinckel93,ivison94}. 
Such broad wings of H$\alpha$
are also found in planetary nebulae including M2-9 \citep{balick89} and IC~4997 \citep{leeh00}.
\citet{vandesteene00}  reported that broad H$\alpha$ wings are found in post AGB stars.
It appears that many active galactic nuclei also exhibit very broad H$\alpha$ wings in their spectra.

Broad H$\alpha$ wings can be formed in a hot tenuous wind \citep[e.g.][]{skopal06}. Many symbiotic stars and
planetary nebulae show P~Cygni type profiles in far UV resonance lines including O~VI$\lambda\lambda$1032 and 1038
and C~IV$\lambda\lambda$1548 and 1551, indicative of the fast outflows from the center.
About 10 percent of quasars show broad absorption trough in their emission lines, which are presumably formed
through resonance scattering in fast outflows from the central supermassive black holes.

Broad H$\alpha$ wings can also be formed through scattering of H$\alpha$ photons with free electrons. \citet{sekeras12} proposed that broad wings
around O~VI$\lambda\lambda$1032, 1038 in symbiotic stars can be well fitted from Thomson scattering in a medium
with Thomson optical depth $\tau_{Th}\sim 0.05-0.8$ and electron temperature $T_e \sim 1.5\times 10^4{\rm\ K}-4\times 10^4{\rm\ K}$
\citep[see also][]{schmid99}.

An interesting astrophysical mechanism giving rise to broad H$\alpha$ wings can be obtained through  Raman scattering
of Ly$\beta$ with atomic hydrogen.
Here, the term `Raman scattering' refers to inelastic scattering of a
far UV photon incident upon a hydrogen atom in the ground state, which finally de-excites into $2s$
state with a reemission of an optical photon. A line photon of O~VI$\lambda$ 1032 is Raman
scattered to become an optical photon with $\lambda=6825{\rm \AA}$ \citep{schmid89} and
a far UV photon near Ly$\beta$ is Raman scattered to appear near H$\alpha$.
The wavelengths of the incident photon and its Raman scattered one,  $\lambda_i$ and $\lambda_o$
are related by
\begin{equation}
\lambda_o^{-1}=\lambda_i^{-1}-\lambda_\alpha^{-1}
\end{equation}
where $\lambda_\alpha$ is the wavelength of Ly$\alpha$.
The inelasticity of scattering results in significant profile broadening by a factor of $\lambda_o/\lambda_i$,
which is about 6 in the case of Ly$\beta$-H$\alpha$ scattering.
The operation of Raman scattering to form H$\alpha$ wings requires a sizable number of Ly$\beta$ photons
from an emission nebula that subsequently enter a neutral region

In Fig.~\ref{raman}, we show a schematic illustration for formation of
H$\alpha$ wings through Raman scattering of Ly$\beta$. 
In this geometry, the Ly$\beta$-H$\alpha$ emission region is
surrounded by a neutral hemispheric region, where
Raman scattering of Ly$\beta$ takes place.
The Ly$\beta$-H$\alpha$ emission source located at the center is the same as
the emission nebula illustrated in the right panel of Fig.~\ref{scheme}. 
%for which we consider two values $N_{HI}=10^{19} \rm cm^{-2}$ and $10^{20} \rm cm^{-2}$ and.
The temperature and the optical depth of the emission source are taken 
to be $T=5 \times 10^4 \rm \, K$ and $\tau_\alpha=1$, respectively.
The hemispheric neutral region is characterized by the column density
$N_{HI} = 10^{19}\, \rm cm^{-2}$.  The bottom
three panels in Fig.~\ref{raman} show resultant wing profiles.
It is noticeable that the HWHM of H$\alpha$ wings in the bottom left panel 
is $125\, \rm km\,s^{-1}$ broader than the HWHM of Ly$\beta$ directly emergent from the source.

In Fig.~\ref{raman_wing}, we present the line profiles of  H$\alpha$ core and wing
parts, formed in the environment described in Fig.~\ref{raman}.
We consider two values of the temperature of Ly$\beta$-H$\alpha$ emission region
 $T=3\times 10^4\, \rm K$ and $5 \times 10^4\, \rm K$. We also vary
the optical depth $\tau_\alpha = 1-100$, and the column density $N_{HI}=10^{19-20} \rm\, cm^{-2}$.
In the case of $N_{HI} = 10^{19} \rm\, cm^{-2}$, single broad wing profiles are obtained.
However, as we increase  $N_{HI}=10^{20}\rm\, cm^{-2}$ and $\tau_\alpha > 5$,
the wing profiles tend to exhibit broad double peak. For example,
the peak separation of broad H$\alpha$ wings is $530\, \rm km\,s^{-1}$ with $\tau_\alpha=10$, $T=5 \times 10^4 \, \rm K$ and $N_{HI} = 10^{20} \rm\, cm^{-2}$.
However, if there is any bulk motion $>30{\rm\ km\ s^{-1}}$ in the Ly$\beta$-H$\alpha$
emission region, the emergent profiles convolved with the bulk motion
may appear as single broad wings.

\cite{arrieta03} showed that the wing profiles are excellently fitted using a function $(\lambda-\lambda_{H\alpha})^{-2}$
, which is consistent with Raman scattering cross section around Ly$\beta$
\citep[e.g][]{lee00, nussbaumer89, schmid89}.
In this section, we present the parameter space consisting of $\tau_\alpha$ and $T$
in which  Ly$\beta$ is emergent sufficiently to account for H$\alpha$ wings typically observed in symbiotic stars.

%%%%%%%%%%%%%%%%%%%%%%%%%%%%%%%%%%%%%%%%%%%%%%%%%%%%%%%%%%%%%%%%%%%%%%%%%%%%%%%

\section{Summary and discussion}

In this work we have used a Monte Carlo code to treat the radiative
transfer of Ly$\beta$-H$\alpha$ line photons in an optically thick spherical
emission region.  When the electron temperature of the emission region is lower than
20,000 K, escape of Ly$\beta$ is quite insignificant due to small population
of hydrogen atoms in $n=2$ levels. However, it is found that
for temperature $\ge 3\times 10^4 {\rm\ K}$ the emergent Ly$\beta$ flux becomes
comparable to that of H$\alpha$ playing an important role as a coolant
of the emission region. As the line center optical depth of H$\alpha$
increases, the emergent line profile of H$\alpha$ becomes broader.
broader.

Symbiotic stars
are known to exhibit strong H$\alpha$ emission with
broad wings and absorption trough blueward of its line center.
 Raman scattering of Ly$\beta$ by atomic hydrogen results in profile
broadening by a factor $6563/1025=6.4$, leading to formation of broad wings
around H$\alpha$. The current work lends strong support to the proposal that
broad H$\alpha$ wings found in many symbiotic stars have their origin in Raman
scattering of Ly$\beta$ emergent in the H$\alpha$ thick emission nebula
around the white dwarf.

The $2s$ level population can be significantly enhanced due to Raman scattering
of far UV radiation. This is because $2s$ level is mainly depopulated by
 continuum two-photon decay, which is characterized by the decay time
of about 8 seconds.  In particular, strong O~VI resonance doublet lines
are incident on H~I region to be Raman scattered leaving the H atom in the $2s$
state. This effect may be important in the blueward dip of H$\alpha$ profiles
observed in many symbiotic stars.

This work is also expected to shed some light on the interpretation
of the Balmer decrement that deviates from
the case B recombination theory in symbiotic stars
\citep[e.g.][]{schwank97}. In particular, the case B recombination
dictates the flux ratio of $F_{H\alpha}/F_{H\beta}=2.86$, and many
symbiotic stars exhibit stronger H$\alpha$ than this generic value.
Escape of higher Lyman series photons will produce complicated flux ratios of Balmer fluxes deviating from
the case B theory.

In an H$\alpha$ thick medium, escape in the form of Ly$\beta$ may reduce the H$\alpha$
flux from the case B value. However, the medium is expected to be also optically thick to H$\beta$,
escape of H$\beta$ can be achieved by changing its identity
as Pa$\alpha$. This implies that a nebula with significant $2s$ population is expected
to exhibit  $F_{H\alpha}/F_{H\beta}$ ratio different than the case B value
of 2.86. A more refined radiative transfer study incorporating Paschen and higher Lyman and Balmer transitions
is required to provide a more satisfactory answer.

Hydrogen emission lines are an important tool to study the cosmic star formation history
and Ly$\alpha$ is particularly useful to probe the galactic environment associated with
star formation activities in the early universe with $z$ higher than 2. Being a prototypical resonance line, 
Ly$\alpha$ photons suffer enormously large number of scatterings before escape leading to formation
of characteristic P~Cygni profiles or profiles suppressed in the blue part
\citep[e.g.,][]{ahn03,dijkstra14}. 
Recent studies show that many
Lyman alpha emitters are surrounded by a neutral halo, where Ly$\alpha$ photons appear to be resonantly
scattered. 

One good example of Ly$\beta$ emission can be found
%The Ly$\beta$ emission breaking case B recombination is exhibited in quasar spectra.
in quasars, where Ly$\beta$ and O~VI form a broad composite emission line.
\cite{laor94,laor95} have analyzed UV spectra of quasars with redshifts in the range 
$0.165 \le z \le 2.06$ to investigate the properties of emission lines
including Ly$\alpha \,1215+$N~V 1240 and Ly$\beta \,1025 +$ O~VI 1034. They deconvolved the composite
broad emission lines to find the flux ratios of Ly$\beta /$O~VI $=0.34\pm0.26$
and Ly$\beta /$Ly$\alpha=0.059\pm0.04$.
\cite{vandenberk01} carried out similar analyses using spectra of quasars 
in an extended range of redshift $0.044 \le z \le 4.789$ from the {\it Sloan Digital Sky Survey}.
The flux ratios found in their study are (Ly$\beta+$O~VI)$/$Ly$\alpha=0.096$ and H$\alpha/$Ly$\alpha=0.31$.

\cite{cabot16} investigated Ly$\alpha$, C~IV and He~II emission lines in Lyman $\alpha$ blobs,
pointing out the presence of emission nebulae with temperature $\sim 2 \times 10^4$ for Ly$\alpha$ and $\sim 10^5{\rm\ K}$ for C~IV and He~II. In this environment,
Ly$\beta$ photons also escape from the hot nebular region and may suffer Raman scattering with hydrogen
atoms that may reside possibly in the neighboring neutral halo. In this case 
we may predict that H$\alpha$ will be characterized by very broad wings that may exceed the kinematic
speed associated with the circumgalactic medium. 
IR spectroscopy using a space telescope such as {\it James Webb Space Telescope} can be performed to detect the broad wings
around Balmer emission lines.

%%% ACKNOWLEDGMENTS (IF ANY) %%%%%%%%%%%%%%%%%%%%%%%%%%%%%%%%%%%%%%%%

\acknowledgments

We are very grateful to an anonymous referee for the
constructive comments, which improved the presentation
of the current paper.
This research was supported by the Korea Astronomy and Space Science Institute under the R\&D program(Project No. 2018-1-860-00) supervised by the Ministry of Science and ICT.
The Monte Carlo calculation was performed by using the PC-cluster Polaris in KASI.
Seok-Jun Chang is also particularly grateful to Dr. Jongsoo Kim for his help
to parallize and improve the code in a much more efficient way.

%%% APPENDICES (IF ANY) %%%%%%%%%%%%%%%%%%%%%%%%%%%%%%%%%%%%%%%%%%%%%

%%% CALL LIST OF REFERENCES (natbib STYLE) %%%%%%%%%%%%%%%%%%%%%%%%%%

%%%%%%%%%%%%%%%%%%%%%%%%%%%%%%%%%%%%%%%%%%%%%%%%%%%%%%%%%%%%%%%%%%%%%%%%%%%%%%%


\begin{thebibliography}{}

%%% PUT YOUR REFERENCES HERE %%%%%%%%%%%%%%%%%%%%%%%%%%%%%%%%%%%%%%%%

\bibitem[Adams (1972)]{adams72} Adams, T. F., 1972, The Escape of Resonance-Line Radiation from Extremely Opaque Media, ApJ, 174, 439
\bibitem[Ahn  et al. (2003)]{ahn03} Ahn, S.-H., Lee, H.-W., Lee, H. M., 2003,
P Cygni type Ly$\alpha$ from starburst galaxies,
MNRAS, 340, 863
\bibitem[Ahn \& Lee (2015)]{ahn15} Ahn, S.-H., Lee, H.-W.,2015,
Polarization of Lyman $\alpha$ Emergent from a Thick Slab of Neutral Hydrogen,
JKAS, 48, 195
%\bibitem[Angeloni et al. (2010)]{angeloni10} Angeloni, R. Contini,  M. Ciroi,  S., Rafanelli, P., 2010, 
%The spectral energy distribution of D-type symbiotic stars: the role of dust shells,
%MNRAS, 402, 2075
\bibitem[Arrieta \& Torres-Peimbert (2003)]{arrieta03} Arrieta, A., Torres-Peimbert, S  2003,
Broad H$\alpha$ Wings in Nebulae around Evolved Stars and in Young Planetary Nebulae,
ApJS, 147, 97
\bibitem[Balick (1989)]{balick89}  Balick, B., 1989,  
M2-9 - A planetary nebula with an eruptive nucleus?,
AJ, 97, 476
\bibitem[Cabot et al. (2016)]{cabot16} Cabot, S., H. C., Cen, R., Zheng, Z., 
C IV and He II line emission of Lyman $\alpha$ blobs: powered by shock-heated gas,
2016, MNRAS, 462, 1076
\bibitem[Chang et al. (2015)]{chang15} Chang, S.-J., Heo, J.-E., Di Mille, F.,
Angeloni, R., Palma, T., Lee, H.-W., 
Formation of Raman Scattering Wings around H alpha, H beta, and Pa alpha in Active Galactic Nuclei,
2015, ApJ, 814, 98
\bibitem[Dijkstra (2014)]{dijkstra14} Dijkstra, M., 
Ly$\alpha$ Emitting Galaxies as a Probe of Reionisation,
2014, PASA, 31, 40
\bibitem[Gnat \& Sternberg (2007)]{gnat07} Gnat, O., Sternberg, A., 
Time-dependent Ionization in Radiatively Cooling Gas,
2007, ApJS, 168, 213
\bibitem[Gronke et al. (2016)]{gronke16} Gronke, Ma., Dijkstra, M., McCourt, M., Oh, S. P., 
From Mirrors to Windows: Lyman-alpha Radiative Transfer in a Very Clumpy Medium,
2016, ApJL, 833L, 26
\bibitem[Godon et al. (2012)]{godon12} Godon, P., Sion, E. M.; Levay, K., Linnell, A. P., Szkody, P.,
An Online Catalog of Cataclysmic Variable Spectra from the Far-Ultraviolet Spectroscopic Explorer,
2012, ApJS, 203, 29G 
%\bibitem[Henize \& McLaughlin (1951)]{henize51} Henize, K. G., McLaughlin, S. B., 
%A Note on the Spectrum of RR Telescopii,
%1951, ApJ, 114, 163
\bibitem[Ivison et al. (1994)]{ivison94}  Ivison, R. J., Bode, M. F., Meaburn, J., 
An atlas of high resolution line profiles of symbiotic stars. II. Echelle spectroscopy of northern sky objects,
1994, A\&AS, 103, 201
\bibitem[Kenyon (1986)]{kenyon86} Kenyon, S., 1986, The Symbiotic Stars, 
New York, Cambridge University Press
\bibitem[Laor et al. (1995)]{laor95} Laor, A., Bahcall, J. N., Jannuzi, B. T., Schneider, D. P., Green, R. F., The Ultraviolet Emission Properties of 13 Quasars, 
1995, ApJS, 99, 1
\bibitem[Laor et al. (1994)]{laor94} Laor, A., Bahcall, J. N., Jannuzi, B. T., Schneider, D. P., Green, R. F., Hartig, G. F., 
The ultraviolet emission properties of five low-redshift active galactic nuclei at high signal-to-noise ratio and spectral resolution, 
1994, ApJ, 420, 110
\bibitem[Lee (1994)]{lee94} Lee, H.-W., 
On the Polarization of Resonantly Scattered Emission Lines - Part Two - Polarized Emission from Anisotropically Expanding Clouds
1994, MNRAS, 268, 49
\bibitem[Lee (2000)]{lee00} Lee, H.-W., 
Raman-Scattering Wings of H$\alpha$ in Symbiotic Stars,
2000,  ApJL, 541, 25
\bibitem[Lee (2013)]{lee13} Lee, H.-W., 
Asymmetric Absorption Profiles of Ly$\alpha$ and Ly$\beta$ in Damped Ly$\alpha$ Systems,
2013,  ApJ, 772, 123
\bibitem[Lee \& Hyung (2000)]{leeh00} Lee, H.-W., Hyung, S., 
Broad H$\alpha$ Wing Formation in the Planetary Nebula IC 4997,
2000, ApJL, 530, 49
\bibitem[Netzer (2015)]{netzer15} Netzer, H.,  
Revisiting the Unified Model of Active Galactic Nuclei,
2015, ARA\&A, 53, 365
\bibitem[Neufeld (1990)]{neufeld90} Neufeld, D. A., 
The transfer of resonance-line radiation in static astrophysical media,
1990, ApJ, 350, 216
\bibitem[Nussbaumer et al. (1989)]{nussbaumer89} Nussbaumer, H., Schmid, H. M., Vogel, M.,
Raman scattering as a diagnostic possibility in astrophysics,
1989, A\&A, 211, 27
\bibitem[Nu\~{n}ez et al. (2016)]{nunez16} Nu\~{n}ez, N. E., Nelson, T., Mukai, K., Sokoloski, J. L., Luna, G. J. M., 
Symbiotic Stars in X-Rays. III. Suzaku Observations,
2016, ApJ, 824, 23
\bibitem[Osterbrock (1962)]{osterbrock62} Osterbrock, D.E., 
The Escape of Resonance-Line Radiation from an Optically Thick Nebula,
1962,  ApJ, 135, 195
\bibitem[Ouchi et al. (2010)]{ouchi10} Ouchi, M., Shimasaku, K., Furusawa, H., Saito, T., Yoshida, M., 
Statistics of 207 Ly$\alpha$ Emitters at a Redshift Near 7: Constraints on Reionization and Galaxy Formation Models,
2010,  ApJ, 723, 869
\bibitem[Rybicki \& Lightman (1985)]{rybicki85} Rybicki, G. B. \& Lightman, A. P., 1985,
Radiative Processes in Astrophysics, John Wiley \& Sons, New York
\bibitem[Saslow \& Mills (1969)]{saslow69} Saslow W. M., Mills D. L., 
Raman Scattering by Hydrogenic Systems,
1969, Phys. Rev., 187, 1025
\bibitem[Sakurai (1967)]{sakurai67} Sakurai J. J., 1967, Advanced Quantum Mechanics. Addison-
Wesley, Reading, MA.
\bibitem[Schmid (1989)]{schmid89} Schmid, H.  
dentification of the emission bands at 6830, 7088 A,
1989, A\&A, 211, 31
\bibitem[Schmid et al. (1999)]{schmid99} Schmid, H. et al.,  
ORFEUS spectroscopy of the O BT VI lines in symbiotic stars and the Raman scattering process,
1999, A\&A, 348, 950
\bibitem[Schwank (1997)]{schwank97} Schwank, M., Schmutz, W., Nussbaumer, H.,
Irradiated red giant atmospheres in S-type symbiotic stars,
1997, A\&A, 319, 166
\bibitem[Sekeras \& Skopal (2012)]{sekeras12} Sekeras, M., Skopal, A., 
Electron optical depths and temperatures of symbiotic nebulae from Thomson scattering,
2012, MNRAS, 427, 979
\bibitem[Sion et al. (2017)]{sion17} Sion, E. M.; Godon, P., Mikolajewska, J., Sabra, B., Kolobow, C., 
FUSE Spectroscopy of the Accreting Hot Components in Symbiotic Variables,
AJ, 153, 160
\bibitem[Skopal (2005)]{skopal05} Skopal, A., 
Disentangling the composite continuum of symbiotic binaries. I. S-type systems,
2005, A\&A, 440, 995
\bibitem[Skopal (2006)]{skopal06} Skopal, A., 
Broad H$\alpha$ wings from the optically thin stellar wind of the hot components in symbiotic binaries,
2006, A\&A, 457, 1003
\bibitem[Van de Steene et al. (2000)]{vandesteene00} van de Steene, G. C.; Wood, P. R.; van Hoof, P. A. M., 
H$\alpha$ Emission Line Profiles of Selected Post-AGB Stars,
2000, ASPC, 199, 191
\bibitem[Vanden Berk et al. (2001)]{vandenberk01} Vanden Berk, D. E. et al.,
Composite Quasar Spectra from the Sloan Digital Sky Survey,
2001, Aj, 122, 549 
\bibitem[Van Winckel et al. (1993)]{vanwinckel93} Van Winckel, H.; Duerbeck, H. W.; Schwarz, H. E., 
An Atlas of High Resolution Line Profiles of Symbiotic Stars - Part One - Coudé Echelle Spectrometry of Southern Objects and a Classification System of H$\alpha$ Line Profile,
1993, A\&AS, 102, 401
\bibitem[Yang et al. (2011)]{yang11} Yang, Y., Zabludoff, A., Jahnke, K., Eisenstein, D., Davé, R., 
Gas Kinematics in Ly$\alpha$ Nebulae,
2011,  ApJ, 735, 87


%%% END LIST OF REFERENCES %%%%%%%%%%%%%%%%%%%%%%%%%%%%%%%%%%%%%%%%%%

\end{thebibliography}
\end{document}